\documentclass[a4paper,eqsecnum,nofootinbib]{revtex4}
\usepackage{slashed,bbold,amsmath,amssymb,mathtools,wrapfig} 
\usepackage{color,epsfig}
\usepackage{hyperref}

\newcommand{\nn}{\nonumber}
\newcommand{\beq} {\begin{equation}}
\newcommand{\eeq} {\end{equation}}
\newcommand{\beqa} {\begin{eqnarray}}
\newcommand{\eeqa} {\end{eqnarray}}

\newcommand{\ie}{{\it i.e.}}
\newcommand{\etc}{{\it etc}}
\newcommand{\eg}{{\it e.g.}}

\newcommand{\as}{{\alpha_s}}
\newcommand{\lqcd}{\Lambda_{QCD}}

\newcommand{\la}{\Lambda}

\newcommand{\vphi}{\varphi}

\newcommand{\order}[1]{${\cal O}\left(#1 \right)$}

\newcommand{\eq}[1]{(\ref{#1})}

\newcommand{\gsim}{\gtrsim}

\newcommand{\inv}[1]{\frac{1}{#1}}
\newcommand{\halft}{{\textstyle \frac{1}{2}}}

\newcommand{\sfrac}[2]{{\textstyle\frac{#1}{#2}}}
\newcommand{\ket}[1]{\left\vert{#1}\right\rangle}
\newcommand{\bra}[1]{\langle{#1}\vert}

\newcommand{\acom}[2]{\left\{{#1},{#2}\right\}}

\newcommand{\bs}[1]{\boldsymbol{#1}}

\newcommand{\mA}{\mathcal{A}}
\newcommand{\mB}{\mathcal{B}}

\newcommand{\mM}{\mathcal{M}}

\newcommand{\xv}{{\bs{x}}} 
\newcommand{\yv}{{\bs{y}}}

\newcommand{\pv}{{\bs{p}}}
\newcommand{\kv}{{\bs{k}}}
\newcommand{\qv}{{\bs{q}}}

\newcommand{\Pv}{{\bs{P}}}
\newcommand{\Av}{{\bs{A}}}

\newcommand{\gv}{\bs{\gamma}}

\newcommand{\gz}{\gamma^0}

\newcommand{\nv}{\bs{\nabla}}

\begin{document}

\title{Born Level Bound States\footnote{Talk presented at the Light Cone 2016 Conference, 5-8 September 2016 in Lisbon, Portugal. Version to be published in Few-Body Systems.}}

\author{Paul Hoyer}
\affiliation{Department of Physics, POB 64, FIN-00014 University of Helsinki, Finland}

\begin{abstract} \parskip 2mm
Bound state poles in the $S$-matrix of perturbative QED are generated by the {\em divergence} of the expansion in $\alpha$. The perturbative corrections are necessarily singular when expanding around free, \order{\alpha^0} $in$ and $out$ states that have no overlap with finite-sized atomic wave functions. Nevertheless, measurables such as binding energies do have well-behaved expansions in powers of $\alpha$ (and $\log\alpha$).

It is desirable to formulate the concept of ``lowest order'' for gauge theory bound states such that higher order corrections vanish in the $\alpha \to 0$ limit. This may allow to determine a lowest order term for QCD hadrons which incorporates essential features such as confinement and chiral symmetry breaking, and thus can serve as the starting point of a useful perturbative expansion.

I discuss a ``Born'' (no loop, lowest order in $\hbar$) approximation. Born level states are bound by gauge fields which satisfy the classical field equations. Gauss' law determines a distinct field $A^0(\xv)$ for each instantaneous position of the charges. A Poincar\'e covariant boundary condition for the gluon field leads to a confining potential for $q\bar q$ and $qqq$ states. In frames where the bound state is in motion the classical gauge field is obtained by a Lorentz boost of the rest frame field.
\end{abstract}

\maketitle

\vspace{-.5cm}

\parindent 0cm
\vspace{-.2cm}

\section{Introduction} \label{intro}

The standard perturbative expansion of the $S$-matrix, defined by the master formula \eq{smatrix} below, is the basis for analytic approaches to the Standard Model. However, Eq. \eq{smatrix} is not valid for {\em bound} states in $i$ and $f$, since they have zero overlap with the free $in$ and $out$ states. Even a first approximation to a bound state (such as that given by the Schr\"odinger equation for atoms) necessarily involves all powers of the coupling. This raises the question whether there is a physically motivated first approximation for bound states, around which a convergent power series in the coupling may be developed. In this talk I discuss \order{\hbar^0} ``Born states''. In analogy to standard tree diagrams the bound state Born terms have no loop contributions. I refer to my lecture notes \cite{Hoyer:2014gna} and \cite{Hoyer:2016aew} for a more detailed treatment.

Section \ref{smatrixsec} concerns the $\hbar$ expansion and the need to modify \eq{smatrix} so that it applies to bound states. Section \ref{born} verifies that the Born term concept gives the standard Schr\"odinger approximation for QED atoms. The (equal time) Hamiltonian formulation furthermore allows to consider atomic states with arbitrary CM momenta. In section \ref{diracfock} I discuss the states of an electron that is relativistically bound by an external potential. The Dirac wave functions determine not only the valence electron distribution, but also those of the $e^+e^-$ pairs. In section \ref{confpot} I study whether QCD Born states can serve as a first approximation for hadrons. Confinement requires to consider homogeneous solutions of Gauss' law for which the instantaneous gluon field $A_a^0(\xv\to\infty)\neq 0$. In solutions which maintain Poincar\'e symmetry each quark color component of the hadron is bound by a linear potential. An external observer does not see the confining $A_a^0$ field of color singlet hadrons since it vanishes when averaged over quark colors. I discuss some properties of the meson solutions in section \ref{messol}, and give a brief outlook in section \ref{summary}.


\section{The perturbative $S$-matrix with bound $in$ and $out$ states} \label{smatrixsec}

Perturbation theory is our main analytic tool in studies of physical gauge theories. Its success in describing QED+EW and (hard) QCD scattering amplitudes has established the Standard Model. The expression for the $S$-matrix in the Interaction Picture (IP) is
\beq\label{smatrix}
S_{fi}={}_{out}\bra{f,\,t\to \infty}\Big\{ {\rm T}\exp\Big[-i\int_{-\infty}^\infty dt\,H_I(t)\Big]\Big\}\ket{i,\,t\to -\infty}_{in}
\eeq
The $in$ and $out$ states are eigenstates of the free, \order{g^0} Hamiltonian $H_0$. The expansion of the exponential in powers of the coupling $g$ ($H_I \propto g$) defines the perturbative corrections. The eigenstates of $H_0$ evolve into eigenstates of the full Hamiltonian $H$ during their evolution from the asymptotic times $t = \pm\infty$. The perturbative expansion is thus formally exact provided that there is a non-vanishing overlap between the $in$ and $out$ states and the physical $i$ and $f$ states of the scattering amplitude.

As it stands, the expression \eq{smatrix} for $S_{fi}$ is inappropriate when the initial or final state involves a (stable) bound state. The infinite separations of free particles (wave packets) means that the $in$ and $out$ states have a vanishing overlap with finite-sized bound states $B$: 
\begin{align}
\bra{B,t=-\infty}i,t=-\infty\rangle_{in} = \bra{B,t=-\infty}U^\dag(t)U(t)\ket{i,t=-\infty}_{in} = 0
\end{align}
By definition, bound states are eigenstates of the full $H$ and thus stationary in time. Hence the overlap of $\ket{B,t}$ will vanish with any state $U(t)\ket{i,t=-\infty}_{in}$ that the $in$ state evolves into. Equivalently, no finite order Feynman diagram can have a bound state pole. 

Inspection of the Feynman diagrams contributing to $e^+e^- \to e^+e^-$ shows 
(see, \eg, sect. II of \cite{Hoyer:2014gna}) 
that diagrams with repeated single photon exchanges (the ``ladder'' diagrams) have a special role in bound state formation. When all momenta are scaled as in atoms ($|\pv| \propto \alpha\, m_e$ in the rest frame) every ladder diagram contributes at the same power of $\alpha$. On the other hand, non-ladders having crossed photons, loop corrections \etc, are suppressed by one or more powers of $\alpha$. 

The sum of all ladders gives (through its divergence) rise to bound state poles. The pole residues show that the (rest frame) wave functions satisfy the Schr\"odinger equation with the classical potential $V(r)=-\alpha/r$. The Schr\"odinger atom thus appears as a first approximation of the physical atom, with perturbative corrections that vanish as $\alpha \to 0$. The ladder sum may be formulated as a Bethe-Salpeter equation for the wave function, valid in any frame and with a single photon exchange kernel. The perturbative corrections can then be systematically added through loop corrections to the kernel and propagators. In the rest frame the Non-Relativistic QED (NRQED) expansion in powers of $|\pv|/m_e$ is very efficient \cite{Kinoshita}. In either case, the perturbative corrections are applied to a first approximation that is guessed, or obtained from a divergent expansion. 

While the above approach is successful for QED atoms it fails to describe confinement and chiral symmetry breaking in QCD hadrons. This could be due to perturbation theory being inapplicable. However, some features of hadrons are strikingly similar to atoms, such as the spectra of heavy quarkonia. It therefore seems worthwhile to investigate the alternative possibility that $\as(0)$ does allow a perturbative expansion. The failure to describe color confinement would be due to the quarks and gluons being infinitely separated in the $in$ and $out$ states of \eq{smatrix}. 

As remarked previously, the Schr\"odinger atom is bound by a {\em classical} potential. This indicates that it is the lowest order term in an $\hbar$ expansion of the exact QED bound state. Recall that a Green function of a (bosonic) field $\vphi$ has the functional integral expression 
\beq\label{green}
G(x_1,\ldots,x_n) = \int [d\vphi]\,\vphi(x_1)\ldots\vphi(x_n)\,e^{iS[\vphi]/\hbar}
\eeq
In the limit $\hbar\to 0$ the (classical) field configurations for which the action $S[\vphi]$ is stationary give the leading contribution. Contributions of \order{\hbar} correspond to a sum over field configurations, or equivalently, integrals over loop momenta in Feynman diagrams. By identifying the Schr\"odinger atom with a Born approximation we avoid the divergent sum of ladder diagrams.

It may seem surprising that an all-orders sum in $\alpha$ can correspond to lowest order in $\hbar$. In scattering amplitudes there is a one-to-one correspondence between the powers of $\alpha$ and $\hbar$ (\ie, the number of loops). However, \order{\hbar^0} (tree) diagrams with $n$ vertices are of \order{g^n}: The power of $g$, but not of $\hbar$, increases with the number of external legs. 
Ladder diagrams are (approximated as) convolutions of \order{\alpha\,\hbar^0} single photon exchange amplitudes. Their geometric sum is non-polynomial in $\alpha$ while remaining of lowest order in $\hbar$.

The Interaction Picture of the $S$-matrix \eq{smatrix} is based on separating the full Hamiltonian $H$ into its free $H_0$ and interacting $H_{int}$ parts, 
\begin{align} \label{intpict}
H = H_0+H_{int}
\end{align}
The usefulness of the IP rests on having analytic (plane wave) solutions of the eigenstates of $H_0$, and on their similarity with the asymptotic scattering states. 
Infrared singularities arise because the scattering amplitudes of charged particles are not gauge invariant. In QED the IR singularities can be removed by associating each charged particle with an infinite number of soft photons \cite{Kulish:1970ut}.

The constituents of (stable) bound states remain close to each other at all times. In order to allow physical Positronium atoms in the initial and final states of the $S$-matrix, their Born terms (Schr\"odinger atoms) should be included in the asymptotic $in$ and $out$ states. $H_0$ must then include the classical $A^0$ field generated by the electron (at $\xv_1$) and positron (at $\xv_2$),
\begin{align}
-\nv_\xv^2 A^0(\xv;\xv_1,\xv_2) &= e\big[\delta(\xv-\xv_1)-\delta(\xv-\xv_2)\big] \label{gausslaw0} \\[2mm]
A^0(\xv;\xv_1,\xv_2) &= \frac{e}{4\pi}\left(\inv{|\xv-\xv_1|}-\inv{|\xv-\xv_2|}\right) \label{qedfield}
\end{align}
$A^0$ is determined instantaneously by the positions of the charges due to the absence of a $\partial_t A^0$ term in the QED Lagrangian. Each position of the charges corresponds to a distinct $A^0(\xv;\xv_1,\xv_2)$. In the rest frame of the atom the spatial, propagating components $\Av$ of the gauge field give contributions of higher order in $\alpha$. Loop effects of \order{\hbar} are generated by $H_I$ and treated perturbatively, with the charged particles propagating in the classical field. Infrared singularities are absent since (neutral) atoms decouple from soft photons. Because the corrections to the Schr\"odinger atom vanish in the $\alpha \to 0$ limit the perturbative expansion thus defined is expected to converge.

The modification of the Interaction Picture envisaged here requires a more precise definition. Here I summarize results obtained in \cite{Hoyer:2014gna} and \cite{Hoyer:2016aew} at the Born level, neglecting the effects of $H_I$.

\section{Hamiltonian formulation of Born level (Schr\"odinger) atoms} \label{born}

A Born level Positronium state in the rest frame may be expressed in terms of the 
electron and positron creation operators (evaluated at any given time $t$),  
\beq\label{nrwf2}
\ket{n,t}=\int \frac{d\kv}{(2\pi)^3}\, \phi_n(\kv)\,b^\dag(\kv,\lambda)\,d^\dag(-\kv,\lambda')\ket{0}
\eeq
where $n$ labels the state and the Schr\"odinger wave function $\phi_n(\kv)$ is independent of the $e^-,\,e^+$ helicities $\lambda,\lambda'$. In terms of the $4\times 4$ wave function 
\beq\label{nrwf}
\Phi_n^{\alpha\beta}(\kv) \equiv \prescript{}{\alpha}{\Big{[}}\gamma^0 u(\kv,\lambda)\Big]\, \Big[v^\dag(-\kv,\lambda')\Big]_\beta\, \phi_n(\kv)
\eeq
the state may in coordinate space be expressed in terms of the electron field $\psi(t,\xv)$,
\beq\label{bstate}
\ket{n,t}=\int d\xv_1 d\xv_2 \,\bar\psi_\alpha(t,\xv_1)\Phi_n^{\alpha\beta}(\xv_1 -\xv_2)\psi_\beta(t,\xv_2)\ket{0}
\eeq
The bound state should be an eigenstate of the QED Hamiltonian,
\begin{align}
H(t) &= \int d\xv\,\Big\{\bar\psi(t,\xv)\big[-i\nv\cdot\gv+m_e+e\gz A^0(\xv)\big]\psi(t,\xv)+{\mathcal H}_{field}\Big\} \label{qedham} \\[2mm]
H(t)\ket{n,t} &= (2m_e+E_n^b)\ket{n,t} \label{bsecond}
\end{align}
where $E_n^b<0$ is the binding energy. In the non-relativistic limit no pairs are produced, $b^\dag d^\dag\ket{n,t} \to 0$. The field energy ${\mathcal H}_{field}$ of the classical field \eq{qedfield} turns out  
(see Eq. (2.18) of \cite{Hoyer:2016aew})  
to cancel half of the contribution from the fermion interaction term $\psi^\dag(t,\xv)\,eA^0(\xv)\psi(t,\xv)$. Its effect can thus be taken into account by a factor $\halft$ in front of this term.

Having determined the Hamiltonian we may impose the stationarity condition \eq{bsecond}. Using the canonical relations $\acom{\psi^\dag_\alpha(t,\xv)}{\psi_\beta(t,\yv)}=\delta_{\alpha,\beta}\,\delta^3(\xv-\yv)$ it gives the Schr\"odinger equation for the wave function $\phi_n$ in \eq{nrwf2}, with potential $V(\xv_1-\xv_2)=-\alpha/|\xv_1-\xv_2|$, as expected.
 
A similar analysis can be carried out for a Positronium state with any momentum $P=(P^0,\Pv)$,
\beq\label{bstateP}
\ket{n,\Pv,t}=\int d\xv_1 d\xv_2 \, e^{i\Pv\cdot(\xv_1 +\xv_2)/2}\,\bar\psi(t,\xv_1)\Phi_n^{(\Pv)}(\xv_1 -\xv_2)\psi(t,\xv_2)\ket{0}
\eeq
Due to the Hamiltonian framework the fields are evaluated at equal time in any frame. Since the definition of time is frame-dependent, Lorentz covariance is not explicit, and emerges as a dynamical consequence of the Poincar\'e invariance of the action. It is thus non-trivial that 
\begin{align}
H(t)\ket{n,\Pv,t} &= \sqrt{\Pv^2+(2m+E_n^b)^2}\,\ket{n,\Pv,t} \label{bsecondP}
\end{align}
This relation is verified (in the limit $\alpha \ll 1$) when the classical gauge field in $H(t)$ is related to the  rest frame field \eq{qedfield} by a standard boost. The Positronium wave function Lorentz contracts similarly to rods in classical relativity,
see sect. II.4 of \cite{Hoyer:2016aew}.  

\section{Fock states of the Dirac wave function} \label{diracfock}

An intriguing aspect of hadrons is that their quantum numbers are determined solely by their valence constituents ($q\bar q$ or $qqq$). The sea quarks and gluons seen in DIS do not enrich the spectrum. This is possible due to the relativistic binding, and as such can be studied  in the Dirac framework of an electron bound in a strong external field $A^\mu(\xv)$. The Dirac wave function is characterized by the quantum numbers of a single electron, but the state it describes must, due to `$Z$-diagrams' and to account for the Klein paradox, contain multiple $e^+e^-$ pairs.

The positive and negative energy solutions of the Dirac equation $(E_n, \bar E_n >0)$,
\begin{align}
\big(-i\nv\cdot\gv+m+e\slashed{A}\big)\Phi_n(\xv) &= E_n\gz\Phi_n(\xv) \nn \\[2mm]
\big(-i\nv\cdot\gv+m+e\slashed{A}\big)\bar\Phi_n(\xv) &= -\bar E_n\gz\bar\Phi_n(\xv) \label{dir2}
\end{align}
determine positive energy eigenstates of the Dirac Hamiltonian operator $H_D$,
\beq\label{ham1}
H_D=\int d\xv\,\bar\psi(\xv)\big[-i\nv\cdot\gv+m_e+e\slashed{A}\big]\psi(\xv)
\eeq
The eigenstates $\ket{n}$ and $\ket{\bar n}$ of $H_D$ with eigenvalues $E_n$ and $\bar E_n$ are given in sect. III of \cite{Hoyer:2016aew},  
\begin{align}
&\ket{n} = \int d\xv\,\psi^\dag_\alpha(\xv)\Phi_{n\,\alpha}(\xv)\ket{\Omega}
&\ket{\bar n} = \int d\xv\,{\bar\Phi}_{n\,\alpha}^\dag(\xv)\psi_\alpha(\xv)\ket{\Omega} 
\end{align}
The ground state $\ket{\Omega}$ is a superposition of Fock states with any number of $e^+e^-$ pairs, 
\beq\label{vac2}
\ket{\Omega} = N_0\exp\Big[-b_q^\dag \big(B^{-1}\big)_{qm}D_{mr}d_r^\dag\Big]\ket{0}
\eeq 
where a sum over the 3-momenta and helicities $q,r$ and over the state labels $m$ is understood. The matrices $B$ and $D$ are given by the Dirac wave functions \eq{dir2} and the free spinors $u,v$,
\begin{align}
B_{mq}= \Phi^\dag_m(\qv)u(\qv,\lambda) \hspace{2cm} D_{mr}= \Phi^\dag_m(-\bs{r})v(\bs{r},\lambda)
\end{align}
The vacuum state satisfies $H_D\ket{\Omega}=0$. In the weak field limit $\ket{n}$ turns into a single electron and $\ket{\bar n}$ into a single positron state.

The Dirac states allow to understand a curious property of the Dirac wave functions that was noted already in 1932: The wave functions cannot be normalized and the energy spectrum is continuous for a linear $A^0$ potential \cite{plesset}. A potential which confines electrons repulses positrons. Hence the states have both a confined and a deconfined (accelerating/decelerating) component. The latter is due to the positrons in the $e^+e^-$ pairs, which (predominantly) appear in the region where $|eA^0(\xv)| \gsim 2m_e$ 
(see Fig. 14 of \cite{Hoyer:2014gna} for an example in $D=1+1$ dimensions). 
The solutions are characterized by a continuous parameter which determines the ratio of $e^+e^-$ pairs to valence $e^-$. The minimum ratio is given by the Schwinger pair production rate \cite{Schwinger:1951nm}.

\section{A confining QCD potential from a boundary condition} \label{confpot}

A primary motivation for the present study is to determine whether the concept of Born term for QED atoms is relevant for QCD hadrons. With no loop contributions the coupling is frozen at a value which may be perturbative, $\alpha_s^{\overline{MS}}/\pi \simeq 0.137$ \cite{Dokshitzer:1998nz}. A confining potential requires a parameter $\lqcd$ with the dimension of mass. This parameter does not appear in the QCD Lagrangian, and can result from the classical gluon field equations only due to a boundary condition. The standard QED solution \eq{qedfield} of Gauss' law \eq{gausslaw0} is obtained with the boundary condition $\lim_{|\xv| \to \infty} A^0(\xv)=0$.

A non-vanishing boundary condition at $|\xv| \to \infty$ generally violates Poincar\'e invariance and generates long-distance effects. However, there is an (apparently unique) acceptable solution for mesons and baryons (see section VI B of \cite{Hoyer:2014gna} for an operator field formulation, and section V C of \cite{Hoyer:2016aew} for the equivalent classical gauge field). Translation invariance requires that the field strength have no spatial dependence and that the states be (global) color singlets. Rotational invariance is preserved when the field is correlated with the positions of the charges, analogously to \eq{qedfield}. A non-vanishing boundary condition 
on the classical field 
can be imposed only on $A^\mu_a$ with $\mu=0$ and $a=3,8$. $\mu=0$ ensures instantaneity and the diagonal color matrices $T_{3,8}^{AB} \propto \delta_{AB}$ conserve the quark colors. This should hold in a gauge where the color structure of the meson wave functions is $\Phi^{AB}\propto \delta^{AB}$ and for baryons $\Phi^{ABC}\propto \epsilon^{ABC}$.

Gauss' law is imposed separately for each spatial position of the quarks as in \eq{gausslaw0}, and for each quark color component. The boundary condition results in a confining potential which for mesons is exactly linear. Since the state is an overall color singlet the field vanishes when summed over colors. Thus an external observer does not see the hadron via its confining field, eliminating long range effects.

The \order{\alpha_s^0\,\hbar^0} confining gauge fields for the $\ket{q_A(\xv_1)\bar q_A(\xv_2)}$ Fock component of a meson are
\begin{align}
\bar\psi_A(\xv_1)\psi_{A}(\xv_2)\ket{0}:\hspace{2cm} \mA_a^0(\xv)= 6\Lambda^2\,\frac{\xv\cdot(\xv_1-\xv_2)}{|\xv_1-\xv_2|} T_a^{AA}  \hspace{2cm} (a=3,8) \label{mfield}
\end{align}
where the constant $\Lambda$ characterizes the boundary condition of this homogeneous solution ($\nv^2 A_a^0(\xv)=0$) of Gauss' law. 
Since $\sum_A T_a^{AA}=0$ a color singlet meson generates no overall color field.

The corresponding color field for the Fock component $\ket{q_1(\xv_1) q_2(\xv_2) q_3(\xv_3)}$ of a baryon is 
\begin{align}
&\psi_1^\dag(\xv_1)\psi_2^\dag(\xv_2)\psi_3^\dag(\xv_3)\ket{0}:\hspace{.3cm} \mA_3^0(\xv)=3\Lambda^2\,\frac{\xv\cdot(\xv_1-\xv_2)}{d(\xv_1,\xv_2,\xv_3)} \hspace{1cm} \mA_8^0(\xv)=\sqrt{3}\Lambda^2\,\frac{\xv\cdot(\xv_1+\xv_2-2\xv_3)}{d(\xv_1,\xv_2,\xv_3)} \nn \\ 
&{\rm where}  \label{bfield} \\
&d(\xv_1,\xv_2,\xv_3)=\inv{\sqrt{2}}\sqrt{(\xv_1-\xv_2)^2+(\xv_2-\xv_3)^2+(\xv_3-\xv_1)^2} \nn
\end{align}
The total field energy $\int d\xv\,\sum_{a=3,8} \big[\nv \mA_a^0(\xv)\big]^2 = 12\la^4\int d\xv$ must be a universal, the same for all Fock states of mesons and baryons. Hence $\la$ is a universal constant, equivalent to $\lqcd$.

Using the classical gauge fields \eq{mfield}, \eq{bfield} in the QCD Hamiltonian the condition 
\beq\label{hmeson}
H_{QCD}\ket{n,\Pv=0} = M_n\ket{n,\Pv=0}
\eeq
determines the bound state equations for the meson and baryon wave functions in the rest frame. Quark pair production (``string breaking'') is of higher order in $1/\sqrt{N_c}$ and can be included iteratively.
The meson bound state equation is, with $\Phi^{AB}(\xv_1-\xv_2) =\delta^{AB}\Phi(\xv_1-\xv_2)/\sqrt{3}$,
\beq\label{mbse}
i\nv\cdot\acom{\gz\gv}{\Phi(\xv)}+m_1\gz\Phi(\xv)-m_2\Phi(\xv)\gz = \big[M-V_\mM(\xv)\big]\Phi(\xv)
\eeq
where $m_{1,2}$ are the (current) quark masses. 
The potential $V_\mM$ arises from the interaction term in $H_{QCD}$. For the $\ket{q_A(\xv_1)\bar q_A(\xv_2)}$ Fock component the color field \eq{mfield} contributes at $\xv=\xv_1$ and $\xv=\xv_2$ (with opposite signs). The field energy subtracts half of the quark interaction energy as in QED \eq{qedham} (see section V D of \cite{Hoyer:2016aew} for details),
\beq\label{mpot}
V_\mM(\xv_1-\xv_2) = \halft g\sum_{a=3,8}T_a^{AA}\big[\mA_a^0(\xv_1)- \mA_a^0(\xv_2)\big] = g\la^2\,|\xv_1-\xv_2|
\eeq
The potential is seen to be independent of the quark color $A$ using the identity $\sum_a T_a^{AB}T_a^{CD}= \halft\delta^{AD}\delta^{BC}-\sfrac{1}{6}\delta^{AB}\delta^{CD}$.

The baryon bound state equation is analogous to \eq{mbse}, with the potential
\beq\label{bpot}
V_\mB(\xv_1,\xv_2,\xv_3) = \inv{\sqrt{2}}\,g\la^2\,\sqrt{(\xv_1-\xv_2)^2+(\xv_2-\xv_3)^2+(\xv_3-\xv_1)^2}
\eeq
The two gluon field components $a=3,8$ in \eq{bfield} induce confinement in the two relative separations $\xv_1-\xv_2$ and $\xv_1+\xv_2-2\xv_3$ of the three quarks in the baryon. No analogous solution is available for multiquark states such as $qq\bar q\bar q$ and $qqq\,q\bar q$. When two quarks coincide the baryon potential reduces to the meson one, $V_\mB(\xv_1,\xv_2,\xv_2)=V_\mM(\xv_1-\xv_2)$.

\section{Properties of the meson solutions} \label{messol}

The rotational invariance of the meson potential \eq{mpot} allows to express the $4\times 4$ wave function in terms of Dirac matrices, spherical harmonics and radial functions. For $q\bar q$ pairs of the same flavor $(m_1=m_2=m)$ the solutions may be grouped into ``trajectories'' \cite{Geffen:1977bh} on which the parity $\eta_P$ and charge conjugation $\eta_C$ depend on the total angular momentum $J$ as $\eta_P = -\eta_C = (-1)^{J+1}$ (``$\pi$ trajectory''), $\eta_P = \eta_C = (-1)^{J+1}$ (``$a_1$ trajectory'') and $\eta_P = \eta_C = (-1)^{J}$ (``$\rho$ trajectory''). Despite the relativistic dynamics there are no states with quantum numbers that would be exotic in the quark model.

The wave function of a state of mass $M$ is in general singular at $M-V(r)=0$. States with regular wave functions have a discrete spectrum. For $m=0$ the Regge trajectories are linear, with evenly spaced daughter trajectories as in dual models (Figs. 16 and 17 of \cite{Hoyer:2016aew}). Similarly to the Dirac case, the radial wave functions have constant norm at large radii $r$. This allows an interpretation of the $q\bar q$ components with large potential energy $V(r)$ as being dual to hadrons produced via string breaking, as in phenomenological models of quark hadronization. The string breaking amplitudes are of higher order in $1/\sqrt{N_c}$ and determined by the overlap of the hadron states obtained at leading order in $1/\sqrt{N_c}$. Thus $A \to B+C$ is given by $\bra{BC}A\rangle$, which can be expressed in terms of the hadron wave functions $\Phi_A,\,\Phi_B$ and $\Phi_C$
(Eq. (7.1) of \cite{Hoyer:2014gna}). 
The square of this amplitude gives a hadron loop correction to $\Phi_A$, as required by unitarity at \order{\hbar^0}. Meson scattering amplitudes may be similarly evaluated.

Analogously to the weakly bound Positronium states, mesons with CM momentum $\Pv$ are bound by the potential obtained by boosting \eq{mfield}. They have energy eigenvalue $E=\sqrt{M^2+\Pv^2}$ and their wave functions are Lorentz contracted compared to the rest frame ones.

The states based on a chiral invariant ground state appear in parity degenerate pairs as expected. There is a massless meson solution with $J^{PC}=0^{++}$. Since it has vanishing 4-momentum it may mix with the ground state without breaking Poincar\'e invariance. This would cause spontaneous chiral symmetry breaking.

\section{Outlook} \label{summary}

The application of perturbation theory to bound states is subtle: the expansion must be developed around an interacting state  (see, \eg, section II C of \cite{Hoyer:2014gna}). Since the Schr\"odinger atom has no loop contributions it serves as a Born term for bound states, analogously to tree diagrams of scattering amplitudes. The \order{\hbar^0} approximation is not limited to non-relativistic dynamics -- Schr\"odinger atoms can be viewed in any reference frame. Loop corrections can be added once the master formula \eq{smatrix} for the $S$-matrix is generalized to include bound state Born terms in the $in$ and $out$ states.

Conceivably, Born bound states could serve as a first approximation also for QCD hadrons. If radiative gluon contributions are suppressed at low scales $Q$ the coupling $\as(Q^2)$ freezes, enabling a perturbative expansion. The novel features of hadrons, including confinement and chiral symmetry breaking, must then appear already at Born level. 

The QCD confinement scale $\lqcd$ can arise from the Born level field equations only via a boundary condition. The instantaneity of $A^0_a$ allows to consider non-trivial homogeneous solutions of Gauss' law. Physical requirements restrict the solutions to color singlet states bound by a linear potential. The spectrum found this way includes massless states, which may mix with the vacuum, causing chiral symmetry breaking.

The perturbative expansion is a central tool in analytic studies of the Standard Model. This, together with features of the data, motivate careful studies of its applicability even to the hadron spectrum and dynamics.

\end{document}